\documentclass{aa}  
\usepackage[colorlinks, allcolors=blue]{hyperref}
\usepackage{amsmath} 
\usepackage{graphicx}
\usepackage{xcolor}
\usepackage{txfonts}
\usepackage{longtable}
\usepackage{booktabs}
\usepackage{geometry}
\usepackage{datetime}
\usepackage{lscape}
\usepackage{float}
\usepackage{placeins}
\usepackage{amssymb}
\usepackage[normalem]{ulem}
\usepackage{upgreek} 

\begin{document} 

   \title{Two sides of the same spin: tracking optical and radio pulsations of SAX~J1808.4-3658 from outburst to quiescence}

\author{C. Ballocco \inst{1,2}
\and A. Miraval Zanon \inst{3}
\and A. Papitto\inst{1}
\and M. Burgay\inst{4}
\and M. C. Baglio\inst{6}
\and C. Malacaria\inst{1}
\and F. Ambrosino\inst{1}
\and A. Possenti\inst{4}
\and G. Illiano\inst{7,8,6}
\and E. D. Barr\inst{21}
\and R. P. Breton\inst{20}
\and L. Burderi\inst{5}
\and S. Campana\inst{6}
\and C. J. Clark\inst{18,19}
\and F. Coti Zelati\inst{7,8,6}
\and D. de Martino\inst{15}
\and T. Di Salvo\inst{16}
\and M. Kramer\inst{21,20}
\and R. La Placa\inst{1}
\and F. Lewis\inst{13,14}
\and A. Marino\inst{9,10,7,8,11}
\and D. M. Russell\inst{17}
\and A. Sanna\inst{5}
\and L. Stella\inst{1}
\and B. Stappers\inst{20}
\and J. D. Turner\inst{1}
\and M. Cecconi\inst{12}
\and A. Ghedina\inst{12}
}

\institute{INAF - Osservatorio Astronomico di Roma, Via Frascati 33, I-00078 Monte Porzio Catone (RM), Italy\\
\email{caterina.ballocco@inaf.it} 
\and Dipartimento di Fisica, Sapienza Università di Roma, Piazzale Aldo Moro 5, I-00185 Rome, Italy
\and ASI - Agenzia Spaziale Italiana, Via del Politecnico snc, I-00133 Rome, Italy 
\and INAF - Osservatorio Astronomico di Cagliari, Via della Scienza 5, I-09047 Selargius (CA), Italy
\and Dipartimento di Fisica, Università degli Studi di Cagliari, SP Monserrato-Sestu km 0.7, I-09042 Monserrato, Italy
\and INAF – Osservatorio Astronomico di Brera, Via Bianchi 46, I-23807, Merate (LC), Italy
\and Institute of Space Sciences (ICE, CSIC), Campus UAB, Carrer de Can Magrans s/n, E-08193 Barcelona, Spain
\and Institut d’Estudis Espacials de Catalunya (IEEC), E-08860 Castelldefels (Barcelona), Spain 
\and Departamento de Física, Universidad de Santiago de Chile (USACH), Av. Víctor Jara 3493, Estación Central, Chile
\and Center for Interdisciplinary Research in Astrophysics and Space Sciences (CIRAS), Universidad de Santiago de Chile
\and INAF/IASF Palermo, via Ugo La Malfa 153, I-90146 - Palermo, Italy
\and Fundación Galileo Galilei-INAF, Rambla José Ana Fernández Pérez 7, E-38712 Breña Baja (Santa Cruz de Tenerife), Spain
\and Faulkes Telescope Project, Cardiff, Wales, UK 
\and The Schools’ Observatory, Astrophysics Research Institute, Liverpool John Moores University, 146 Brownlow Hill, Liverpool L3 5RF, UK  
\and INAF - Osservatorio Astronomico di Capodimonte, Salita Moiariello 16, I-80131 Naples, Italy
\and Dipartimento di Fisica e Chimica – Emilio Segrè, Università di Palermo, Via Archirafi 36, 90123 Palermo, Italy
\and Center for Astrophysics and Space Science (CASS), New York University Abu Dhabi, PO Box 129188, Abu Dhabi, UAE
\and Max Planck Institute for Gravitational Physics (Albert Einstein Institute), D-30167 Hannover, Germany
\and Leibniz Universität Hannover, D-30167 Hannover, Germany
\and Jodrell Bank Centre for Astrophysics, Department of Physics and Astronomy, The University of Manchester, Manchester M13 9PL, UK
\and Max-Planck-Institut für Radioastronomie, Auf dem Hügel 69, D-53121 Bonn, Germany
}

\date{Received September 15, 1996; accepted March 16, 1997}

\authorrunning{Ballocco et al.}
\titlerunning{Tracking optical and radio pulsations of SAX~J1808.4$-$3658} 

\abstract 
{
High-time resolution optical and UV observations have recently revealed unexpectedly bright coherent pulsations from accreting millisecond X-ray pulsars. Their luminosity exceeds that expected from thermal emission from neutron-star hotspots or X-ray reflection by the accretion disk. Characterizing optical pulsations over a wide range of accretion rates through coordinated multi-wavelength observations is expected to shed light on their origin.
We present a timing study of the accreting millisecond pulsar SAX J1808.4-3658 during its 2022 outburst and quiescent state, using high-time-resolution optical observations with TNG/SiFAP2, simultaneous NICER X-ray data, optical monitoring from LCO, and MeerKAT radio observations in quiescence. 
During the outburst, we detect coherent optical pulsations, with pulsed luminosities of $\sim1.3\times10^{30}\,\rm{erg\,s^{-1}}$.
Combined with previous detections, our results suggest that the optical pulsed amplitude increases as the X-ray flux decreases. The X-ray and optical pulsed luminosities appear to show opposite trends with X-ray luminosity, suggesting different responses of the two pulsed components to changes in the accretion rate.
We also measure X-ray-to-optical delays of $0.88 \pm 0.04\,\text{ms}$ and $0.51 \pm 0.03\,\text{ms}$ on two different outburst observations, indicating that the relative timing between the two signals is not constant. Together with the spectral energy distribution, these results support different emission mechanisms for the pulsed radiation in the two bands, with non-thermal emission from relativistic particles producing optical pulsations and channeled accretion onto the neutron-star surface powering X-ray pulsations.
During quiescence, we detect no coherent optical pulsations, obtaining a 95\% confidence upper limit of $L^{\text{V, UL}}_{\text{pulsed}}<2.2\times 10^{29}\, \mathrm{erg\,s^{-1}}$ on the optical pulsed luminosity, corresponding to an optical efficiency comparable to those of spider radio millisecond pulsars. Simultaneous MeerKAT observations reveal no radio pulsations down to $\sim10\,\upmu\text{Jy}$, significantly improving previous constraints, but not excluding the presence of an active rotation-powered pulsar whose radio emission may be absorbed by intra-binary material.
}

   
   \maketitle
\section{Introduction}
Accreting millisecond pulsars (AMSPs) represent an evolutionary link between low-mass X-ray binaries (LMXBs) and rotation-powered millisecond pulsars (MSPs; see, e.g., \citealt{PatrunoWatts_2021, DiSalvoSanna_2022} for recent reviews). 
In the recycling scenario, old slow-spinning neutron stars (NSs) are spun up to millisecond spin periods through the accretion of matter and angular momentum from a low-mass companion star \citep{Alpar_1982, BHATTACHARYA_1991}. When accretion stops, these NSs might switch on as radio MSPs, although this has been observed only in one case so far \citep{Papitto_2013}.
The discovery of the first AMSP SAX~J1808.4-3658 (hereafter SAX~J1808) confirmed that mass accretion in LMXBs can spin up a NS to millisecond spin periods \citep{wijnands_1998}. Coherent X-ray pulsations at $\sim 401\, \text{Hz}$ are detected during outbursts, likely produced by the magnetospheric channeling of accreted matter from the companion star toward the NS magnetic poles.
In the two decades since this discovery, the number of known AMSPs has increased significantly, with 26 systems identified to date (see, e.g., \citealt{DiSalvoSanna_2022} and references therein and the most recent discovery in \citealt{Ng_2024, Molkov_2024}). Among them, SAX~J1808 remains the most extensively studied source, having exhibited 11 outbursts since its discovery.
During outbursts, the source reaches X-ray luminosities of $\sim10^{36-37} \,\mathrm{erg\,s^{-1}}$ \citep{gilfanov_1998}, before returning to a quiescence level of $\sim 5 \times 10^{31}\,\mathrm{erg\, s^{-1}}$ \citep{Stella_2000, Campana_2004, Heinke_2007, Hartman_2008}. 

SAX~J1808 was also the first discovered AMSP to exhibit coherent optical and UV pulsations  \citep{Ambrosino_2021}. First detected during the 2019 outburst, optical/UV pulsations were later confirmed during the 2022 outburst \citep{Ballocco_2026}. 
The high luminosity of both optical and UV pulsations ($L_{\rm pulsed(opt)}\approx3\times10^{31}$\, erg\, s$^{-1}$, $L_{\rm pulsed (UV)}\approx2\times10^{32}$\, erg\, s$^{-1}$) challenge standard emission models involving accretion onto the NS's surface.
Only a few other NS binary systems have shown coherent optical and/or UV pulsations, most notably the transitional MSP (tMSP) PSR~J1023+0038 \citep[e.g.,][]{Ambrosino_2017, Papitto_2019, Jaodand_2021, Miraval_2022}. Evidence for optical pulsations has also been reported from the rotation-powered redback MSP PSR~J2339$-$0533, albeit at a much lower statistical significance \citep{Papitto_2025}.
In PSR~J1023+0038, optical and X-ray pulsations exhibit a nearly constant phase offset of approximately 200~$\upmu$s, and are detected when the source is in a sub-luminous disk state at an average X-ray luminosity of $L_{\rm{X}} \simeq 4\times 10^{33}\, \text{erg s}^{-1}$ \citep{Illiano_2023b, Conforti_2026}. Optical pulsations in this source are interpreted in terms of synchrotron emission produced at the boundary between the accretion flow and the relativistic wind of a rotation-powered pulsar \citep{Papitto_2019, Veledina_2019, Baglio_2023, Baglio_2024}. 
In contrast, SAX~J1808 exhibited a half-cycle phase shift between optical and X-ray pulsations \citep{Ambrosino_2021}, which corresponds to a $\sim 1$~ms time lag, suggesting that they originate from different mechanisms at distinct locations. Considering also that bright X-ray pulsations demonstrated that mass accretion onto the NS surface was actually taking place, the origin of the pulsed optical and UV emission in SAX~J1808 remains unclear.

Indirect observational evidence suggests that SAX~J1808 becomes active as a rotation-powered pulsar in its quiescent state.
First, the 11 outbursts observed so far from this source have allowed a detailed study of the pulsar’s long-term spin-frequency evolution. The coherent timing of the pulsations has revealed a constant secular spin-down derivative of $\sim 10^{-15}\, \text{Hz/s}$  \citep{DiSalvo_2008, Hartman_2008, Hartman_2009, Patruno_2012, Bult_2020, Illiano_2023}. This energy loss is consistent with the expected dipole radiation from a rotating magnetic field of $\sim 10^8$~Gauss, as estimated from the mass accretion rate at which X-ray pulsations are observed \citep{Psaltis_1999}.
Further evidence comes from the companion's irradiation in quiescence. The quiescent optical light curve is modulated at the $\sim 2\,\mathrm{hr}$ orbital period of the binary system, with the optical maxima occurring when the donor star is at superior conjunction \citep{Homer_2001, Campana_2004, Deloye_2008, Wang_2009}. Such a modulation typically indicates that the companion's facing side is heated by X-ray irradiation from the accretion flow onto the NS surface \citep{Homer_2001}. 
However, the observed quiescent X-ray flux is insufficient to account for the large amplitude of the optical modulation. The optical excess could be explained if the NS turns on as a rotation-powered radio pulsar during quiescence, producing a pulsar wind strong enough to irradiate the inner face of the companion star that explains the optical excess \citep{Burderi_2003, Campana_2004}.
A possible Fermi-LAT $\gamma$-ray counterpart to SAX~J1808 has also been reported by \citet{Wilhelmi_2016}. If confirmed, this would further support the hypothesis that the system hosts an active rotation-powered pulsar during quiescence.

Despite these indications for an active rotation-powered MSP during quiescence, no radio pulsations have been detected to date from SAX~J1808 in quiescence \citep{Homer_2001, Burderi_2003, Campana_2004, Iacolina_2010, Patruno_2017}, nor from the rest of the other AMSPs for which radio pulsations have been searched (see, for a summary, \citealt{Marino_2026}, Table 5). In the quiescent state, the pulsar wind might prevent accretion and expel matter from the system, creating a surrounding plasma that can significantly attenuate radio signals \citep{Burgay_2003}. 
The non-detection of radio pulsations in SAX~J1808 may thus be explained by the presence of material surrounding the system, which can absorb the radio emission through free-free processes \citep{Burderi_2001, Burderi_2003, DiSalvo_2008, Campana_1998}. 
This contrasts with the tMSP IGR~J18245–2452, which has been observed to switch between accretion-powered and rotation-powered states with detectable radio pulsations in the latter phase \citep{Papitto_2013b}. 

To investigate the properties and nature of optical pulsations seen during an accretion outburst, and to test whether the source switches on as a rotation-powered MSP in quiescence, we present here optical observations performed with the fast photometer SiFAP2 obtained both during the 2022 outburst and in quiescence. To follow the evolution of the 2022 outburst, we also analyzed X-ray observations obtained with NICER and optical monitoring data from the Las Cumbres Observatory (LCO) network.
In addition, we analyzed MeerKAT radio observations obtained in quiescence to search for coherent pulsations.
The significantly higher sensitivity of MeerKAT compared to past radio campaigns on this source allows us to probe much deeper flux density limits and therefore place tighter constraints on the emission geometry and mechanisms in this system.

\section{Observations and data reduction}
Table~\ref{table:obs} summarizes the observations presented and analyzed in this paper. In this section, we describe in detail the procedures used to reduce and analyze each data set.
For convenience, we summarize in Table~\ref{tab:par} the main physical and orbital parameters of SAX~J1808 adopted throughout this work.

\begin{table*}
\renewcommand{\arraystretch}{1.2}
\centering
\caption{Summary of the observations of SAX~J1808.}  
\label{table:obs}      
                                      
\begin{tabular}{c c c c c c}          
\hline                      
Telescope & Instrument/Detector & Mode & Grism/Band & Start time & Net exposure \\
 & & & &[UTC] & [ks]\\
\hline
\multicolumn{6}{c}{Outburst state} \\
\hline
TNG & SiFAP2 & Fast timing & 320--900\,nm & August 26, 2022 21:04:16 & 7.8\\
TNG & SiFAP2 & Fast timing & 320--900\,nm & August 27, 2022 20:56:00 & 8.3\\
NICER & XTI & Fast timing & 0.3--10\,keV & August 26, 2022 01:31:40 & 8.7\\
NICER & XTI & Fast timing & 0.3--10\,keV & August 27, 2022 00:46:00 & 9.7\\
LCO & & Imaging & V, B, R, g', r', i', z & August 21, 2022 & \\
\hline
\multicolumn{6}{c}{Quiescent state} \\
\hline
TNG & SiFAP2 & Fast timing & 320--900\,nm & June 6, 2021 00:55:00 & 4.8\\
TNG & SiFAP2 & Fast timing & 320--900\,nm & June 7, 2021 02:52:00 & 5.4\\
TNG & SiFAP2 & Fast timing & 320--900\,nm & May 29, 2022 01:17:00 & 12.4\\
TNG & SiFAP2 & Fast timing & 320--900\,nm & May 20, 2023 01:41:30 & 11.4\\
TNG & SiFAP2 & Fast timing & 320--900\,nm & May 21, 2023 01:49:30 & 10.8\\
TNG & SiFAP2 & Fast timing & 320--900\,nm & May 25, 2025 01:44:00 & 10.8\\
MeerKAT & APSUSE & SearchMode & L-band & June 5, 2021 01:03:33 & 3.5\\
MeerKAT & APSUSE & SearchMode & L-band & June 6, 2021 01:18:50 & 3.5\\
MeerKAT & APSUSE & SearchMode & L-band & June 7, 2021 01:03:30 & 3.5\\
MeerKAT & APSUSE & SearchMode & L-band & May 29, 2022 01:46:48 & 3.4\\
MeerKAT & APSUSE & SearchMode & L-band & May 30, 2022 02:06:19 & 3.5\\
MeerKAT & APSUSE & SearchMode & L-band & May 31, 2022 01:09:46 & 3.5\\
\hline
\end{tabular}
\end{table*}

\begin{table}
\renewcommand{\arraystretch}{1.2}
\centering
\caption{Physical and orbital parameters of SAX~J1808 adopted throughout this work.}
\begin{tabular}{lccc}
\toprule
Parameter & Value & Reference \\
\midrule
R.A. (J2000) & $\rm{18^{h}08^{m}27^{s}.647(7)}$ & \cite{Bult_2020}\\
decl.(J2000) & $-36^{\circ}58'43''.90(25)$ & \cite{Bult_2020}\\
$\nu$ [Hz]  & 400.975209557(50) & \cite{Illiano_2023} \\
$P_{\rm{orb}}$ [s] & 7249.1600(13) & \cite{Illiano_2023} \\
$a \sin{i}/c$ [lt-s]  & 0.0628033(57) & \cite{Illiano_2023}\\
$T_{\rm{asc}}$ [MJD] &59810.6179996(17) & \cite{Illiano_2023} \\
distance [kpc] & 2.7 & \cite{galloway_2024} \\
$N_{\rm{H}}$ [$\text{cm}^{-2}$] & $1.46(12) \times 10^{21}$ & \cite{DiSalvo_2019}\\
$A_V$ [mag] & 0.51(4) & This work \\
\bottomrule
\\
\end{tabular}
\label{tab:par}
\end{table}

\subsection{NICER} \label{sec:nicer}
\begin{figure*}
\centering
\resizebox{\hsize}{!}{\includegraphics[clip=true]{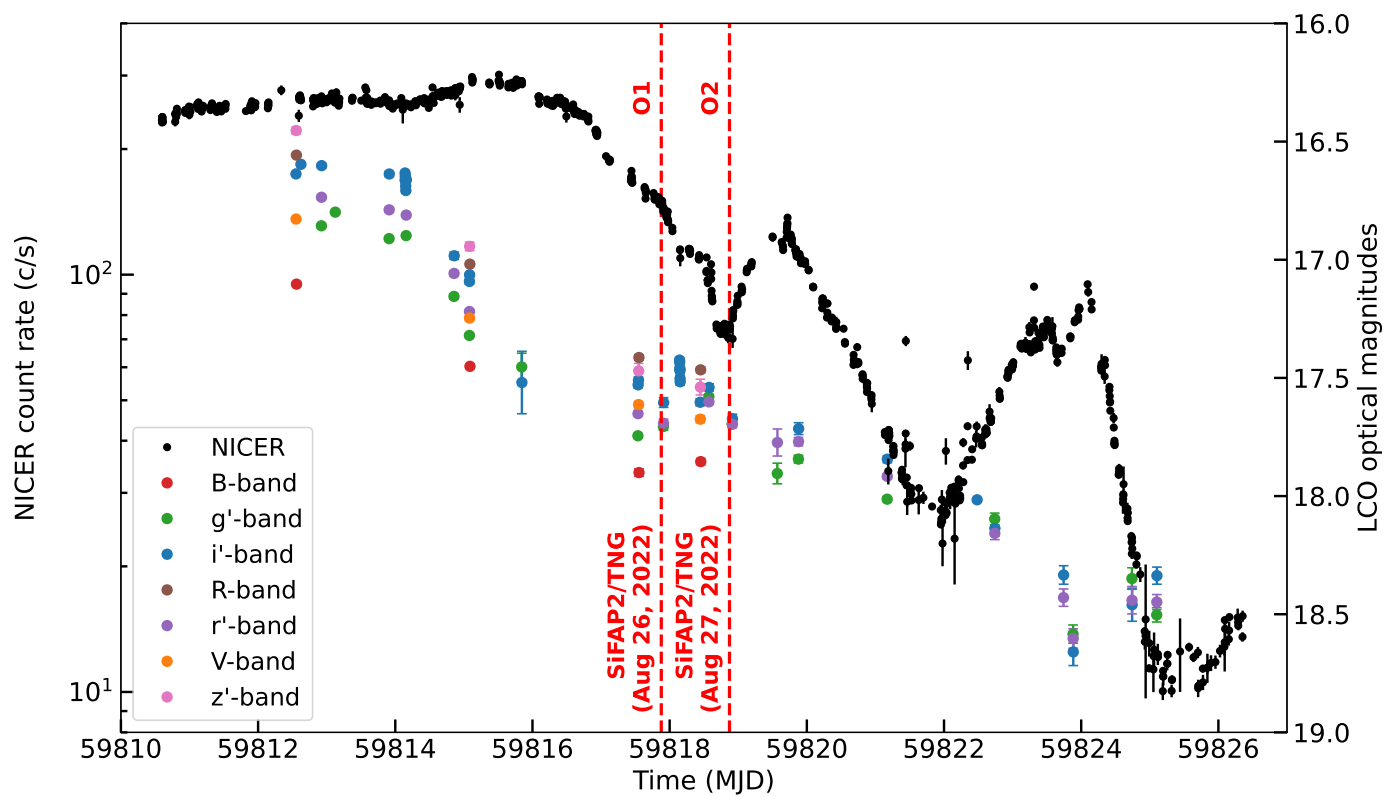}}
\caption{
Multi-wavelength light curve of the 2022 outburst of SAX~J1808. Black points show the NICER light curve in the 0.5--10~keV energy band, using $\mathrm{200\,s}$ bins. Red dashed lines indicate the epochs of the TNG/SiFAP2 observations on 2022 August 26 and 27, respectively O1 and O2. Colored points represent the optical magnitudes (right y-axis) in various bands from the LCO monitoring, uncorrected for interstellar extinction.} 
\label{fig:lctot}
\end{figure*}

The X-ray Timing Instrument (XTI) on board the Neutron star Interior Composition Explorer \citep[NICER;][]{Gendreau_2012} monitored the entire 2022 outburst of SAX~J1808 from August 19 until October 31 \citep{Illiano_2023}. 
In particular, NICER observed SAX~J1808 simultaneously with optical SiFAP2 observations on 2022 August 26-27 (ObsIDs 5574010104 and 5574010105; PI: A. Papitto).
Figure~\ref{fig:lctot} shows the X-ray light curve of the 2022 outburst.
We processed and reduced the data using the standard NICER analysis software \texttt{NICERDAS}, part of HEASoft (v\,6.35.2). Clean and calibrated event files were produced using the \texttt{nicerl2} task.
Using the \texttt{BARYCORR} tool, we corrected the photon arrival times to the Solar System barycenter using the JPL DE405 ephemerides and the source coordinates R.A.~(J2000) = $\rm{18^{h}08^{m}27^{s}.647(7)}$ and decl.(J2000) =$-36^{\circ}58'43''.90(25)$ from \citet{Bult_2020}.
We used the clean event files to produce spectral products with the \texttt{nicerl3-spect} task, applying the \texttt{SCORPEON} background model. We grouped the spectra according to the “optimal binning” scheme of \citet{Kaastra_2016}, with a minimum of 25 counts per energy bin.

\subsection{TNG/SiFAP2}\label{sec:sifap}
\begin{figure}
\centering
\resizebox{\hsize}{!}{\includegraphics[clip=true]{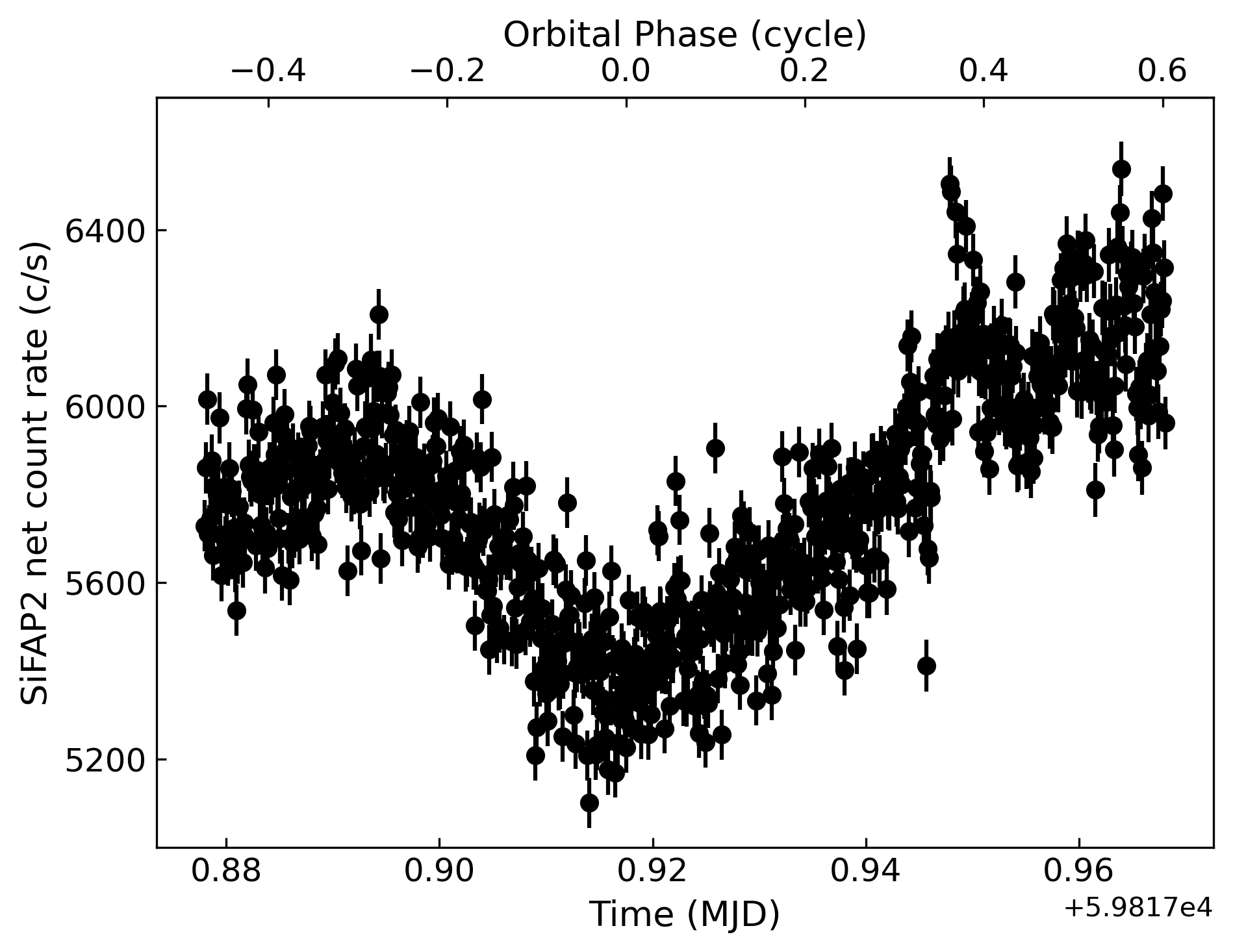}}
\hspace{5mm}
\resizebox{\hsize}{!}{\includegraphics[clip=true]{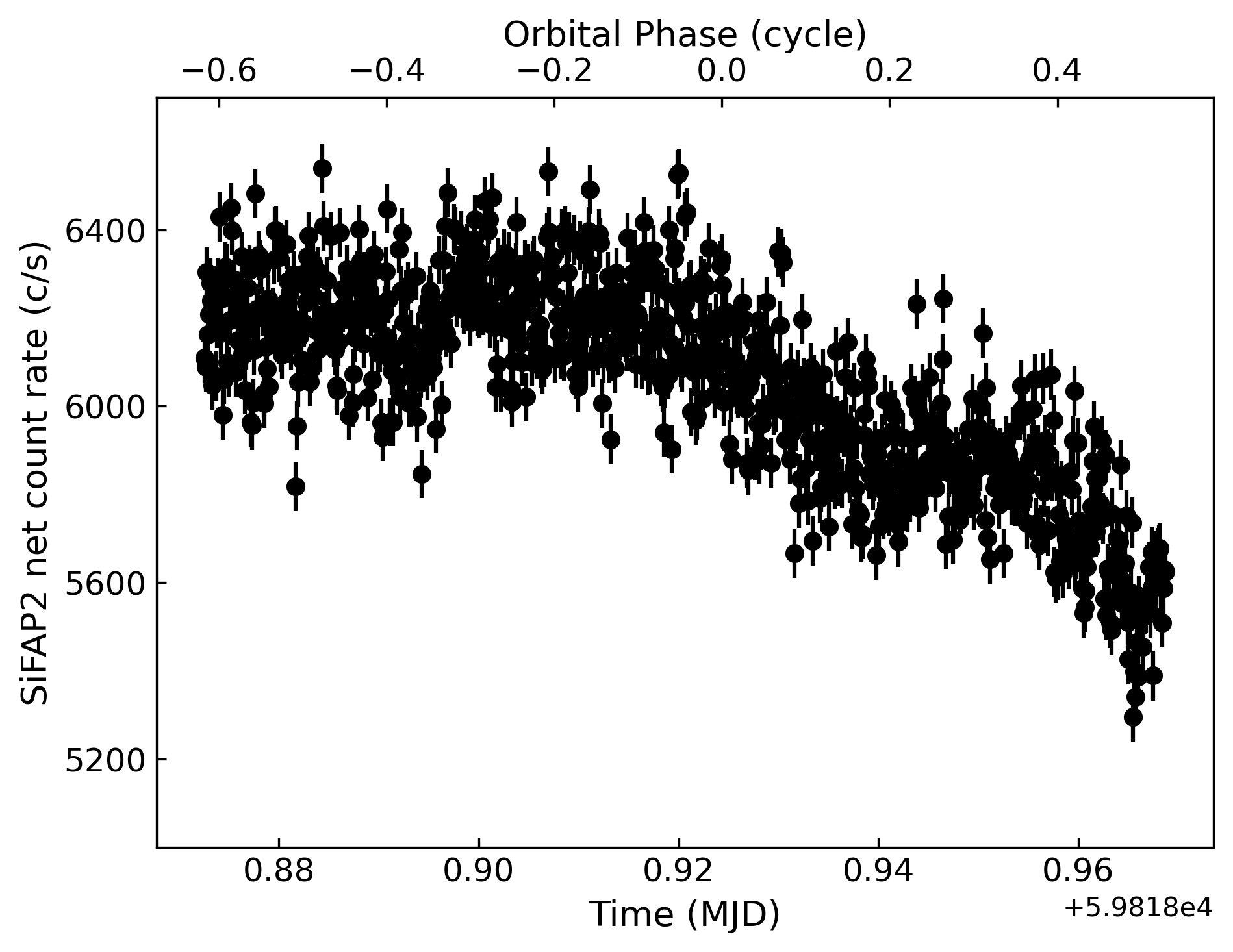}}
\caption{SiFAP2 net light curves of SAX~J1808 during O1 and O2, using 10\,s bins. The observed orbital phase interval is $[-0.47, 0.60]$ for O1 (top panel) and $[-0.61, 0.53]$ for O2 (bottom panel). Error bars represent 1$\sigma$ uncertainties.}
\label{fig:lc_sifap}
\end{figure}

The Silicon Fast Astronomical Photometer \citep[SiFAP2;][]{Ghedina_2018}, mounted at the Telescopio Nazionale Galileo (TNG), observed SAX~J1808 during 2022 outburst for two consecutive nights (red dashed lines in Fig.~\ref{fig:lctot}), on 2022 August 26 for an exposure of 7.8\,ks (hereafter O1, PI: A. Miraval Zanon) and on 2022 August 27 for 8.3\,ks (O2, PI: A. Miraval Zanon).  During quiescent phases of the source, as identified from the absence of X-ray outbursts in the MAXI monitoring light curve, we obtained observations in several epochs between 2021 and 2025 (see Table~\ref{table:obs} for details), for a total exposure of 55.6\,ks.
All observations were carried out using a white filter (320-900\,nm).
The SiFAP2 instrument consists of two independent detectors, each covering a field of view of $7\times7$\,arcsec$^{2}$ and providing an 8\,ns time resolution for individual photon events. During the observations, one detector was centered on the target source. At the same time, the second was simultaneously pointed at a nearby source-free region, located 4.2\,arcmin away, providing an estimate of the sky contribution.
In a subset of the observations, the sky background near the source was also independently estimated by offsetting the telescope pointing by $\sim$10\,arcsec from the target position, for short intervals of $\sim 30$\,s (the so-called nodding).
The net light curves of SAX~J1808 during O1 and O2, shown in Fig.~\ref{fig:lc_sifap}, were obtained by subtracting the background (sky) light curve from the on-source detector light curve.
During O1, we measured an average photon count rate of $\rm{R_{{TNG, O1}}} = (19613 \pm 2)\,\mathrm{c\, s^{-1}}$, with the background contributing approximately 72\% of the total signal ($\mathrm{BKG_{TNG,O1} = (14224 \pm 1)\, c\,s^{-1}}$). For O2, the total average photon count rate was $\rm{R_{{TNG, O2}}} = (18423 \pm 2)\,\mathrm{c\, s^{-1}}$, of which the background accounts for roughly 70\% ($\mathrm{BKG_{TNG,O2} = (12784 \pm 1)\, c\,s^{-1}}$).
We corrected the optical photon arrival time to the Solar System barycenter using the source position derived by \citet{Bult_2020} and the geocentric location of the TNG, along with the JPL DE405 ephemeris.

\subsection{LCO monitoring}
SAX~J1808 was monitored at optical wavelengths with the LCO \citep{Brown2013} network of telescopes as part of an ongoing monitoring campaign of LMXBs coordinated by the Faulkes Telescope Project \citep{Lewis2008}. Observations were obtained in the $B, g', i', R, r', V$ and $z$ bands, with individual exposure times of 100\,s per filter \citep{Baglio2022_Atel}. The data used here cover the 2022 outburst of SAX~J1808, spanning MJD 59812--59825 (2022 August 21 -- September 3). The resulting data set consists of 4, 18, 38, 4, 17, 4, and 4 epochs in the $B, g', i', R, r', V$ and $z$ bands, respectively. The optical imaging data were reduced with the LCO pipeline BANZAI \citep{McCully2018} and subsequently analyzed with the \textit{X-ray Binary New Early Warning System} \citep[XB-NEWS;][]{Russell_2019, Goodwin2020, Saikia2026} pipeline, as described below.

Magnitudes were extracted using multi-aperture photometry (MAP; \citealt{Stetson90}) through XB-NEWS. For each reduced image, XB-NEWS first detects sources using SExtractor \citep{Bertin1996} and solves for the astrometric solution using astrometry.net \citep{Lang2010}, matching against Gaia DR2 positions. If the target is not detected within 1\arcsec\ of its known coordinates at the default threshold, the pipeline re-runs detection at a lower threshold; if still undetected, forced photometry is performed at the target position. Photometry is carried out for all sources using both MAP and fixed-aperture photometry, with aperture radii scaled to multiples of the image PSF full width at half maximum (FWHM). Light curves are then constructed using the DBSCAN clustering algorithm \citep{Ester96}.

Flux calibration is performed by matching sources against an extended version of the ATLAS-REFCAT2 photometric catalogue \citep{Tonry2018}, which incorporates Pan-STARRS DR1 \citep{Chambers2016} and APASS DR10 \citep{Henden2018}. A photometric model is fit to the light curves (in magnitudes) of matched and unmatched sources, including spatially variable zero-point offsets, point spread function (PSF)-based terms, and source-specific mean magnitudes \citep{Bramich2012}; the model is iterated twice with outlier down-weighting to mitigate the effect of intrinsic variability. Colour terms are excluded from the model due to the lack of guaranteed simultaneous multi-band coverage, introducing only small systematic errors (<1--2\%) in the absolute calibration. For the SDSS $i'$ and $r'$-band photometry, XB-NEWS employs Pan-STARRS1 $i_{\rm P1}$ $r_{\rm P1}$ standard magnitudes (AB system), while for the Bessel V-band photometry Johnson V standard magnitudes (Vega system) are used. The fitted photometric model is then applied to calibrate all light curves, including that of the target.
If forced photometry was required for the target, measurements with uncertainties $>0.25\,\rm mag$ were discarded as unreliable.
The resulting multi-band light curves are shown in Fig.~\ref{fig:lctot}.

\subsection{MeerKAT}
Radio observations were acquired with the MeerKAT radio telescope as part of the TRAnsients and PUlsars with MeerKAT (TRAPUM) Large Survey Project's searches for pulsations from gamma-ray sources. MeerKAT is an interferometer array consisting of 64 13.5-m dishes located in the Karoo desert (South Africa). At the time of observing, 60 of the 64 antennas were available and were used to form a coherent tied-array beam at the position of the X-ray source through the GPU-based software beamformer FBFUSE \citep[Filterbanking BeamFormer User Supplied Equipment;][]{Barr2018+FBFUSE, Chen2021+Beamformer}. Data were acquired with the L-band receiver, operating at a central frequency of 1284\,MHz. The 856-MHz bandwidth was split into 4096 frequency channels, and the data were 8-bit sampled every 76.56\,$\upmu$s in 2021 or 38.28\,$\upmu$s in 2022. Data were recorded in filterbank format\footnote{\url{https://sigproc.sourceforge.net/sigproc.pdf}} using the APSUSE (Accelerated Pulsar Search User Supplied Equipment) backend. To speed up processing while maintaining sufficient time resolution to detect a 2.49-ms pulsar, all data were subsequently down-sampled to 76.56\,$\upmu$s off-line. 
Observations were taken during X-ray quiescence in June 5, 6, 7 2021 and in May 29, 30, 31 2022. They were initially scheduled to run simultaneously with the SiFAP2 observations; weather conditions impacted the optical schedule, so that only one radio epoch overlapped the optical observations (see Table~\ref{table:obs}). Each observation lasted approximately 1\,hr (3438 to 3551\,s).

\section{2022 outburst data analysis}
\subsection{Timing analysis}\label{sec:timing}
We performed a timing analysis of the TNG/SiFAP2 observations of SAX~J1808 performed during the decay phase of the 2022 outburst. Using the timing solution derived from NICER data by \citet{Illiano_2023}, we corrected the optical photon arrival times for the binary motion of the pulsar.
To obtain a more accurate measurement of the optical pulse frequency, we performed an epoch folding search using $m=10$ phase bins and a period resolution of $\delta P_{\text{O1,EFS}} = P^2/(2mT_{\rm{obs}}) = 4 \times 10^{-11}\,\text{s}$ for O1 and $\delta P_{\text{O2,EFS}} = 3.7 \times 10^{-11}\,\text{s}$ for O2, where $\mathrm{T_{obs}}$ is the total exposure time of each observation. 
We measured in O1 a maximum chi-squared value of $\mathrm{S_{max,O1}= 98.5}$, corresponding to a best-fitting period of $P_{\text{O1,EFS}} = 2.49391976(7) \times10^{-3} \, \text{s}$, while for O2 we found $\mathrm{S_{max,O2}=194.8}$ and $P_{\text{O2,EFS}} = 2.49391968(4)\times 10^{-3} \,\text{s}$. The $1\sigma$ uncertainty reported was computed following \citet{Leahy_1987}. The measured periods are consistent with the value derived from the X-ray data and reported by \citet{Illiano_2023}. 
We then folded the SiFAP2 observations at the best-fit spin period using 10 phase bins and modeled the resulting optical pulse profiles with a single sinusoidal component.
The background-subtracted root-mean-square (r.m.s.) amplitudes are $A^{\text{rms}}_{\text{O1}} = (0.28\pm 0.03)\%$ and $A^{\text{rms}}_{\text{O2}} = (0.35\pm 0.03)\%$, respectively for the first and second night.
The background contribution was accounted for by applying correction factors of $3.6393(12)$ and $3.2674(9)$ respectively, computed as $\rm{R_{{TNG}}}/(\rm{R_{{TNG}}}-\mathrm{BKG_{TNG}})$, where $\rm{R_{TNG}}$ is the total photon count rate and $\mathrm{BKG}_{\rm{TNG}}$ is the background count rate, as defined in Sect.~\ref{sec:sifap}.

\begin{figure}
\centering
\resizebox{\hsize}{!}{\includegraphics[clip=true]{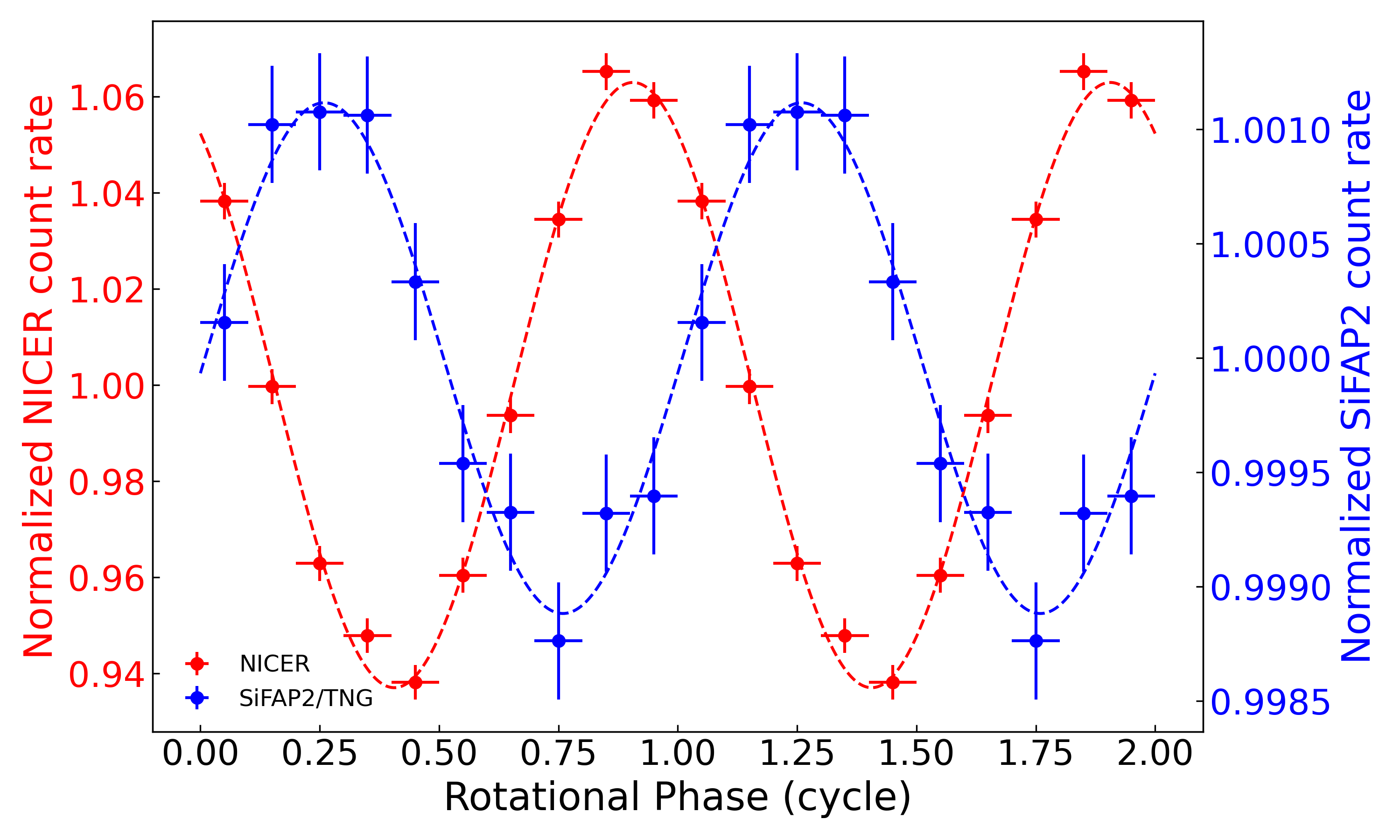}}
\hspace{5mm}
\resizebox{\hsize}{!}{\includegraphics[clip=true]{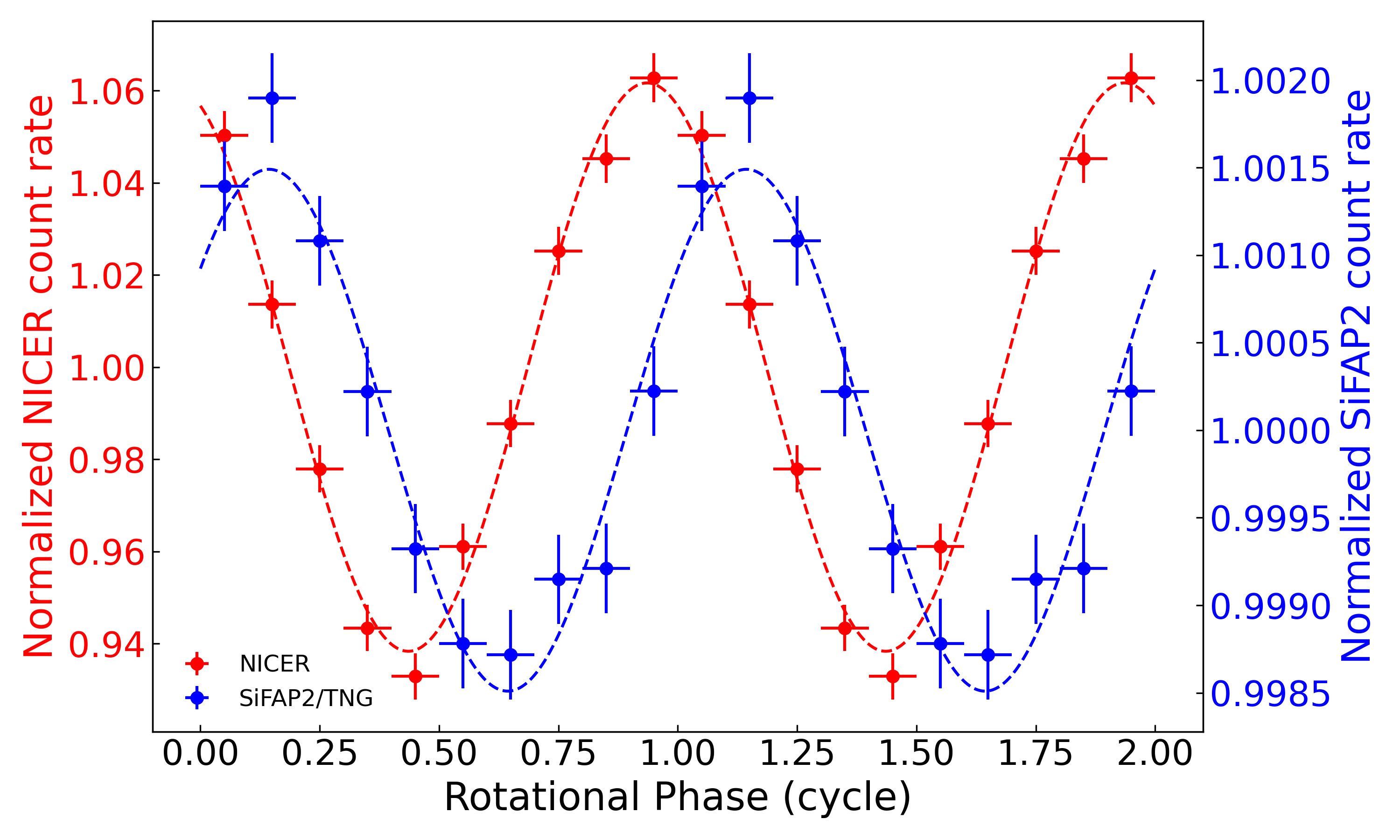}}
\caption{Pulse profiles of SAX~J1808 in the X-ray (red) and optical (blue) bands for O1 (top) and O2 (bottom). The phase shift between the X-ray and optical profiles are $\delta \phi_{O1} = 0.353\pm0.017$ for O1 and $\delta \phi_{O2} = 0.205 \pm 0.013$ for O2, corresponding to time lags of $\delta \tau_{O1}= 0.88 \pm 0.04\, \text{ms}$ and $\delta \tau_{O2}= 0.51 \pm 0.03\, \text{ms}$, respectively. Note the different vertical scales in the two bands.}
\label{fig:lags}
\end{figure}

To compare the X-ray and optical pulsations, we analyzed the NICER observations acquired simultaneously with the SiFAP2 data.
First, we corrected the photon arrival times using the orbital parameters measured by \citet{Illiano_2023}. 
We measured the X-ray pulse background-subtracted r.m.s. amplitude over the time intervals simultaneous with each of the two SiFAP2 observations, obtaining values of $A^{\text{rms}}_{\rm{X1}}=(4.72\pm0.12)$\% and $A^{\text{rms}}_{\rm{X2}}=(4.68\pm0.17)$\%, respectively.
We obtained the pulsed profiles by folding both the SiFAP2 and NICER data into 10 phase bins, using the same spin frequency and a common reference epoch ($\mathrm{T_ {ref}=59817\, MJD}$). This procedure revealed that the optical pulsations lag the X-ray pulsations by phase shifts of $\delta \phi_{O1} = 0.353\pm0.017$ and $\delta \phi_{O2} = 0.205 \pm 0.013$ (see Fig.~\ref{fig:lags}). These values correspond to time lags of $\delta \tau_{O1}= 0.88 \pm 0.04\, \text{ms}$ and $\delta \tau_{O2}= 0.51 \pm 0.03\, \text{ms}$, respectively.
The absolute timing accuracy of SiFAP2 was estimated to be $<60\,\mathrm{\upmu s}$ by \citet{Papitto_2019}, whereas the corresponding value for NICER\footnote{\url{https://heasarc.gsfc.nasa.gov/docs/nicer/mission_guide/}} is $<300\,\text{ns}$. Both values are well below the uncertainty on the measured time lags, indicating a significant variation of the optical/X-ray phase lag between the two observations and also with respect to the value measured in 2019 ($\delta \phi_{2019} = 0.45\pm 0.02$, namely $\delta \tau_{2019}= 1.11 \pm 0.06\, \text{ms}$ \citealt{Ambrosino_2021}).

\subsection{Spectral energy distribution}\label{sec:sed}
We obtained an estimate of the SAX~J1808 flux and luminosity in the optical V-band using data collected from the Faulkes/LCO monitoring of the 2022 outburst (colored points in Fig.~\ref{fig:lctot}). The magnitudes measured were derived using the “X-ray Binary New Early Warning System” \citep[XB-NEWS;][]{Russell_2019}.
We corrected the optical magnitudes for interstellar extinction using the empirical relation for our Galaxy $A_V = N_{\rm{H}}/[(2.87 \pm 0.12) \times 10^{21} \, \text{cm}^{-2}]$ \citep{Foight_2016}, where $N_{\rm{H}} = (1.46\pm0.12) \times 10^{21} \, \text{cm}^{-2}$ \citep{DiSalvo_2019} is the hydrogen column density along the line of sight to SAX~J1808, giving $A_V = 0.51(4)$\,mag. 
The extinction coefficients in the individual photometric bands were then derived from $A_V$ using the parametrized Milky Way extinction law of \citet{gordon23}, evaluated at the average Galactic value of $R_V = 3.1$. For each band, we adopted its effective wavelength (0.4361, 0.5448, and 0.6407~$\upmu$m for Johnson $B$ and $V$ and Cousins $R$, respectively, and 0.4770, 0.6231, 0.7625, and 0.9134~$\upmu$m for the Sloan-like $g'$, $r'$, $i'$, and $z'$ filters used at LCO) and computed the corresponding ratio $A_\lambda/A_V$ from the extinction curve. Scaling these ratios by $A_V = 0.51(4)$~mag, with the uncertainty propagated directly from that on $A_V$, yields $A_B = 0.68(5)$, $A_V = 0.51(4)$, $A_R = 0.42(3)$, $A_{g'} = 0.61(5)$, $A_{r'} = 0.44(3)$, $A_{i'} = 0.33(3)$, and $A_{z'} = 0.24(2)$~mag.

The dereddened V-band magnitudes from the LCO monitoring, obtained within $\sim$10 hours of the SiFAP2 observations, were $17.11(4)\,$mag and $17.16(4)\,$mag for O1 and O2, respectively.
We also estimated the V-band magnitudes from the average SiFAP2 count rates using the instrumental calibration curve\footnote{See \url{https://www.tng.iac.es/instruments/sifap2/}}. The target was observed at a mean relative air-mass of $\sim2.9$ during O1 and O2, which translates into an atmospheric extinction\footnote{\url{https://www.ing.iac.es/Astronomy/observing/manuals/ps/tech_notes/tn031.pdf}} of 0.29(1) mag at 550\,nm.
We obtained values of $17.0(5)\,$mag for O1 and $16.9(5)\,$mag for O2. These are consistent, within SiFAP2 uncertainties, with the magnitudes derived from the LCO monitoring. In the following, we adopt the LCO values, which provide a more precise photometric estimate.
Assuming a constant flux density over the V-band (adopting a FWHM of 88~nm), the corresponding V-band fluxes are $F_{\rm{V,O1}}=(4.72 \pm 0.18)\times 10^{-13}\,\mathrm{erg\, cm^{-2}\, s^{-1}}$ and $F_{\rm{V,O2}}=(4.52 \pm 0.18)\times 10^{-13}\,\mathrm{erg\, cm^{-2}\, s^{-1}}$ for O1 and O2, respectively.
Based on the measured optical pulse amplitude (see Sect.~\ref{sec:timing}) the coherent optical pulsations detected during SiFAP2 observations correspond to a pulsed luminosity of $L_{\rm{pulsed}}^{\rm V,O1} = 0.0028 \, L_{\rm{V,O1}} \approx (1.15\pm0.13) \times 10^{30} \,\text{erg s}^{-1}$ and $L_{\rm{pulsed}}^{\rm V,O2} = 0.0035 \, L_{\rm{V,O2}} \approx (1.38\pm0.13) \times 10^{30} \,\text{erg s}^{-1}$. Unless otherwise stated, all luminosities reported in this work are calculated assuming a source distance of $2.7$~kpc \citep{galloway_2024}.

To estimate the X-ray flux in the 0.5--10~keV energy range at the epochs of the optical TNG/SiFAP2 observations, we performed a spectral fit of the simultaneous NICER data (see Sect.~\ref{sec:nicer}). The spectral analysis was carried out using the \texttt{HEASARC XSPEC} package \citep{Arnaud_1996_XSPEC}, version 12.15.0. Lower energies were excluded due to the dominance of instrumental noise.
We fit the spectra with an absorbed blackbody plus thermal Comptonization model, \texttt{TBabs*(bbodyrad+nthComp)} \citep{Zdziarski_1996, Zycki_1999}. This model is commonly adopted to describe the broadband continuum of AMSPs \citep{Poutanen_2006b, DiSalvoSanna_2022}.
We accounted for interstellar absorption using the \texttt{TBabs} model, adopting the abundances of \citet{Wilms_2000} and the photoelectric cross-sections of \citet{Verner_1996}.
We estimated the 0.5--10~keV X-ray fluxes at the time of SiFAP2 observations, obtaining values of $F_{\rm{X1}}=(4.172\pm0.013)\times 10^{-10}\, \mathrm{erg\,cm^{-2}\,s^{-1}}$ and $F_{\rm{X2}}=(2.145\pm0.010)\times 10^{-10}\, \mathrm{erg\,cm^{-2}\,s^{-1}}$ for O1 and O2, respectively. Using these values and the measured pulse X-ray r.m.s. amplitudes (see Sect.~\ref{sec:timing}), we derived pulsed X-ray luminosities of $L_{\rm{pulsed}}^{\rm X1} = (1.72\pm0.04)\times 10^{34} \, \text{erg s}^{-1}$ and $L_{\rm{pulsed}}^{\rm X2} = (8.8\pm0.3)\times 10^{33} \, \text{erg s}^{-1}$.

\begin{figure}
\centering
\resizebox{\hsize}{!}{\includegraphics[clip=true]{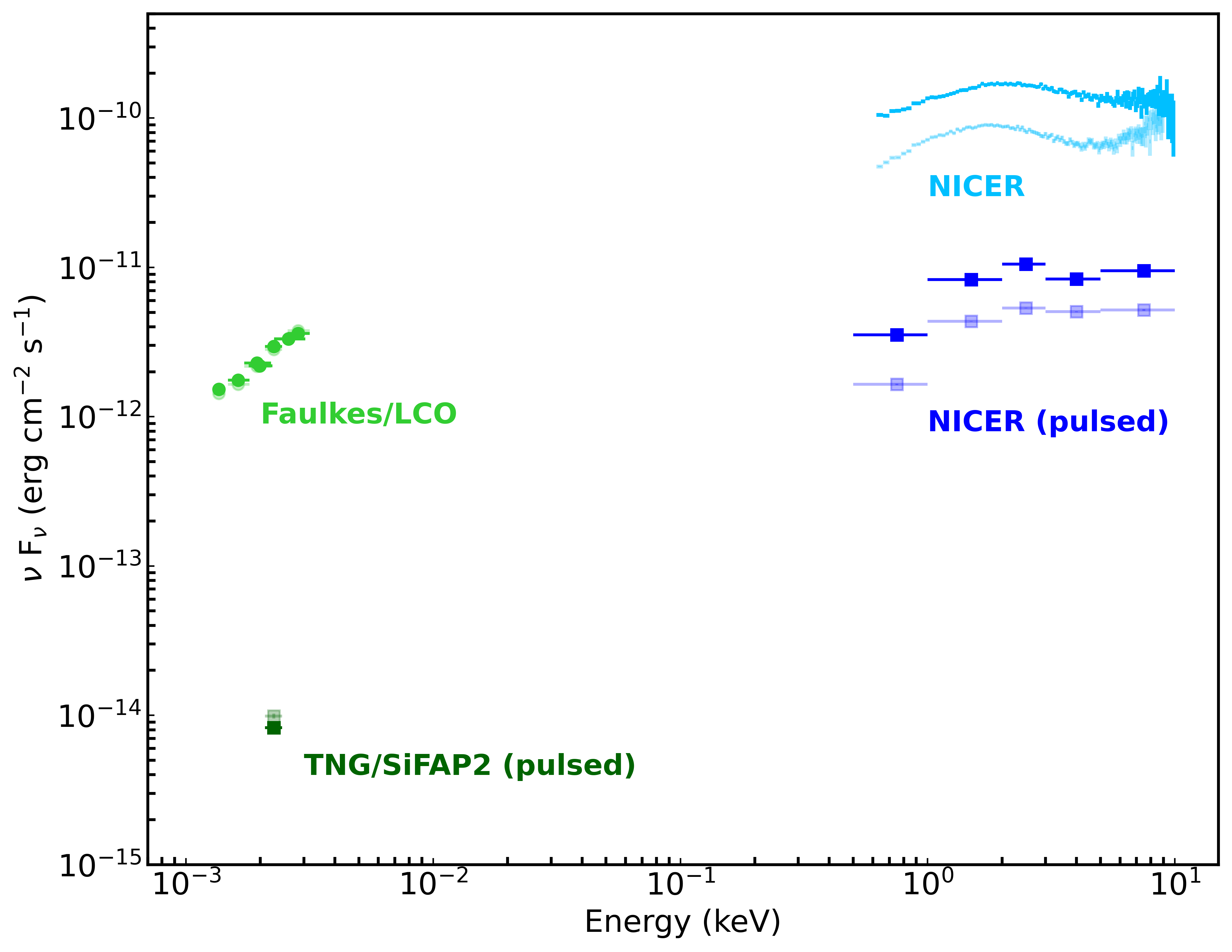}}
\caption{Unabsorbed SED of SAX~J1808 from the optical to the X-ray band obtained from observations performed on 2022 August 26 (fully colored symbols) and 27 (half-transparent symbols). The total X-ray fluxes observed by NICER in the 0.5–10~keV energy band are shown as cyan/light-blue small points. The pulsed X-ray fluxes are shown as dark-blue squares. The optical fluxes derived from the LCO/Faulkes monitoring are plotted as light-green points, while the optical pulsed flux measured with TNG/SiFAP2 in the V-band is shown as dark-green squares. } 
\label{fig:sed}
\end{figure}

Figure~\ref{fig:sed} shows the spectral energy distributions (SED) for the total and pulsed emissions of SAX~J1808 on 2022 August 26 and 27, corrected for interstellar extinction.
To estimate the X-ray pulsed flux, we measured the background-subtracted pulse r.m.s amplitudes in the 0.5--1, 1--2, 2--3, 3--5, and 5--10\,keV energy bands. The pulsed flux in each band was obtained by multiplying the r.m.s. amplitude by the corresponding unabsorbed flux. Finally, these values were converted into $\nu F_{\nu}$ units by scaling them by the ratio between the mid-point energy of each interval and the width of the corresponding energy band.
An analogous procedure was adopted for the optical band: the dereddened pulsed optical fluxes were converted into $\nu F_{\nu}$ units by scaling them by the ratio between the filter central wavelength and its bandwidth.

\section{Quiescence data analysis}
\subsection{Optical upper limit}
To search for optical pulsations during the SiFAP2 observations performed during the quiescent state of SAX~J1808, we first applied the same barycentric corrections described above.
\citet{Illiano_2023} studied the secular evolution of the orbit of SAX~J1808, showing that the modulation of the epoch of passage at the ascending node can be described by a quadratic function plus a sinusoidal term. This allowed us to predict $T_{\rm{asc}}$ at the epoch of our observations.
We first performed a search for the epoch of passage at the ascending node, $T_{\rm{asc}}$, exploring a grid of trial values spaced by $10$\,s \citep[see, e.g.,][]{Caliandro_2012} around the value predicted at the epoch of each observation using Eq.~3 from \citet{Illiano_2023} and spanned the uncertainty of the predicted $T_{\rm{asc}}$. For each value of $T_{\rm{asc}}$, we carried out an epoch-folding search with 32 phase bins, adopting a fixed spin period and orbital parameters derived by \citet{Illiano_2023}. The use of 32 phase bins was motivated by the narrow pulse profiles observed in the radio and $\gamma$-ray bands, as well as by the low-duty-cycle optical pulses detected from rotation-powered MSPs \citep[e.g.,][]{Papitto_2019, Papitto_2025, Malacaria_2026}. The uncertainty on the spin period is well below the resolution required to affect the phase folding over the considered exposure time, ensuring that no additional trials on the spin period are necessary.

To assess the statistical significance of the optical pulsations, we first considered the epoch-folding $\chi^2$ test \citep{Leahy_1983, Leahy_1987}. In the absence of a signal and under the assumption of purely Poissonian noise, the epoch-folding statistic is expected to follow a $\chi^2_{n-1}$ distribution.
However, as shown by \citet{LaPlaca_2026}, SiFAP2 noise follows a generalized Poisson distribution with a dispersion index of $\sim1.15$, rather than the value of 1 that would be expected in the case of purely Poissonian statistics.
We therefore rescaled the observed epoch-folding $\chi^2$ values by this factor to recover the standard $\chi^2_{n-1}$ distribution and evaluate the statistical significance. We set the detection threshold by requiring a probability consistent with a 3.5$\sigma$ confidence level (c.l.) under the null hypothesis of pure counting noise, taking into account the number of trials in the search.

To derive upper limits on the signal strength at the 95\% c.l., we followed the method detailed by \citet{Vaughan_1994} for the analysis of power spectra, based on the formalism originally developed by \citet{Goth_1975}. The latter showed that the power observed in the presence of both signal and noise follows a non-central $\chi^2$ distribution: this is because the total power in a bin stems from the squared modulus of the sum of the signal and noise vectors. In the case of folded profiles, since the total statistic is obtained from the combined squared contributions of the individual phase bins, a similar reasoning can be applied, although made slightly more convoluted by the different variances in each bin (La Placa, Miraval Zanon et al., in prep.).
The resulting upper limits on the signal strength were then converted into upper limits at 95\% c.l. on the r.m.s. amplitude of the signal using the relation $\mathrm{A_{UL}^{rms}=\sqrt{2(\chi^2_{\mathrm{UL}})/N_{\gamma}}}$ \citep{Leahy_1987}, where $N_\gamma$ is the total number of photons.
To take into account the background contribution, we corrected the upper limits using a correction factor defined as $N_{\rm{\gamma(tot)}}/(N_{\rm \gamma(tot)} - N_{\rm \gamma(bkg)})$.
Nodding was not observed during observations in quiescence. In some cases, no simultaneous sky measurements were available, preventing a homogeneous background determination across the data set. Moreover, \citet{Papitto_2025} showed that, even when available, the sky contribution measured far from the source is not a good proxy of the actual background affecting the SiFAP2 target observation.
For this reason, rather than applying the standard background subtraction procedure, we estimated the expected source net count rate from the optical magnitudes measured during quiescence in previous works \citep[see e.g.,][]{Deloye_2008}. Using the SiFAP2 calibration curve, we obtained an expected source net count rate of $\sim150$--250~counts\,s$^{-1}$.
Applying the standard background subtraction procedure instead yielded significantly higher count rates, indicating that the background was likely underestimated in these observations.
Consequently, we estimated the correction factor by constraining the net source count rate ($N_{\rm \gamma(tot)} - N_{\rm \gamma(bkg)}$) to the value expected from the quiescent optical magnitude. Assuming a V-band magnitude of $\sim$21.7~mag \citep{Deloye_2008} and using the SiFAP2 calibration relations, this corresponds to an expected net count rate of $\sim 200\, \mathrm{c\,s^{-1}}$. Individual observations yield upper limits on the optical pulse amplitude ranging down to $\mathrm{A^{rms}_{UL}} < 2.18\%$ in the most constraining dataset.

To derive a single upper limit on the optical pulse amplitude during quiescence for SAX~J1808, we combined all the available observations through an incoherent sum. We summed the maximum $\chi^2$ values obtained from the $T_{\rm asc}$ search for each observation. We then applied the same procedure described above, adopting for the noise a $\chi^2$ distribution with a number of degrees of freedom equal to the sum of those of the individual data sets. 
We thus derived a 95\% c.l. upper limit on the r.m.s. amplitude of the optical pulsations of $\mathrm{A^{rms}_{UL}} < 2.25\%$.
Using the source quiescent $V$-band magnitude, this translates into a 95\% c.l. upper limit on the pulsed optical flux of $F^{\text{V, UL}}_{\text{pulsed}}<2.5\times 10^{-16}\, \mathrm{erg\,cm^{-2}\,s^{-1}}$. This corresponds to an upper limit on the pulsed optical luminosity of $L^{\text{V, UL}}_{\text{pulsed}}<2.2\times 10^{29}\, \mathrm{erg\,s^{-1}}$.

\subsection{Radio upper limit}
After downsampling, MeerKAT radio data processing was performed on the \textit{trg} cluster \citep{MUSCAS_2026} at the INAF Cagliari Observatory (Italy). Data were preliminarily cleaned from radio frequency interferences (RFI) using the \texttt{rfifind} code of the PulsaR Exploration and Search TOolkit PRESTO\footnote{\url{https://github.com/scottransom/presto}} \citep{2011ascl.soft07017R}, with default parameters and an integration time of 2\,s. To correct for the a priori unknown frequency-dependent dispersive delay introduced by the ionized component of the interstellar medium (ISM) in the radio signal, a set of dispersion-corrected time series were created using PRESTO's \texttt{prepsubband} code. The code was run over a grid of 3772 dispersion measures (DMs) ranging from 10.0, a value below which terrestrial RFI would likely become predominant, to 198.55\,pc\,cm$^{-3}$. The maximum DM was chosen to be double the highest value predicted by the ISM models NE2001 \citep{NE2001}, YMW16 \citep{YMW16} and NE2025 \citep{NE2025} for the sky position of SAX~J1808 and for a distance of 2.7\,kpc. The 0.05\,pc\,cm$^{-3}$ DM step adopted was selected using the \texttt{DDplan.py} Python script so that the pulse smearing between two adjacent DMs was at most 10\% larger than the sampling time.  

Each of the 3772 cleaned, de-dispersed time series of each of the 6 observations was searched for radio pulsations in three different ways: i) by folding the time series starting from the X-ray ephemeris and with a varying time of ascending node $T_{\rm{asc}}$, ii) through an FFT-based blind search over the whole observation and over shorter segments, and iii) through a search for single dispersed pulses. 

The folding and $T_{\rm{asc}}$ search were carried out using the PRESTO-based \texttt{spider\_twister}\footnote{\url{https://github.com/alex88ridolfi/SPIDER\_TWISTER}} Python code. Following the same approach used to analyze the quiescent optical data, we searched for orbital phase offsets in the range $-0.005$ to $0.005$. A 10.78\,s step in $T_{\rm{asc}}$ was chosen so that the maximum amplitude of the residual orbital modulation over the observation would be at most 1/4 of the spin period \citep{Ridolfi2016}. The folding was performed using the X-ray ephemerides of \citet{Bult_2020} and, later, of \citet{Illiano_2023}, with 7 trials in $T_{\rm{asc}}$. 

A blind search for periodic, linearly accelerated signals was also run on all the de-dispersed time series using \texttt{pulsar\_miner}\footnote{\url{https://github.com/alex88ridolfi/PULSAR\_MINER}}, a Python pipeline that automatically runs PRESTO routines to search, select, and fold pulsar candidates. For SAX~J1808, we ran it over the full observations and over half-hour, 15-minute, 10-minute and 5-minute segments. The \texttt{accelsearch} code was used with the parameter \texttt{zmax} (which specifies the maximum number of Fourier bins that a binary pulsar signal can drift across during an observation) set to 130, twice the value expected for an object with SAX~J1808 orbital parameters observed for half an hour. This blind, segmented search may be more sensitive than a fully coherent folding in case the signal is visible only in some segment of the orbit and/or the X-ray ephemeris cannot be reliably extrapolated to the epochs of quiescence.

Finally, we also searched for de-dispersed single pulses by running PRESTO's \texttt{single\_pulse\_search.py} on all the DM time series, using a maximum boxcar of 1\,ms and a signal-to-noise (S/N) threshold of 8.

No convincing candidate was found in any of the three searches we performed. To estimate flux density and single-pulse peak flux density upper-limits, we use the modified radiometer equation \citep{1984bens.work..234D}:
\begin{equation}
S_{\rm min} = \beta\, {\rm S/N} \frac{ (T_{\rm sys} + T_{\rm sky}) }{G \sqrt{n_{\rm pol}\,t_{\rm obs}\,{\rm BW}} } \sqrt{\frac{W}{P-W}}  \,\,\, \rm{mJy},
\end{equation}
where $\beta$ is a correction factor due to digitization ($\sim1$ for an 8-bit-sampled time-series), $T_{\rm sys}$ is the system temperature (18\,K for the L-band receiver, plus 4.5\,K of spillover), $T_{\rm sky}$ is the sky temperature (4.4\,K for SAX~J1808, scaled from the 408\,MHz value by \citealt{Haslam1982} to 1.28\,GHz using a spectral index of $-$2.7), G is the telescope gain (2.64\,K\,Jy$^{-1}$ for 60 antennas), $n_{\rm pol}$ is the number of polarizations (2), BW is the bandwidth in MHz (reduced by $\sim18\%$ after RFI excision), $t_{\rm obs}$ is the observing time, $P$ is the pulsar spin period, and $W$ is the pulse width. 

We used a minimum S/N of 6 for the folding+$T_{\rm{asc}}$ search and a value of 8 for the blind and single-pulse searches. For the latter, the peak flux density upper limit can be calculated using the same radiometer equation, removing the spin-period-related part and using the duration of the pulse (in a range from one time sample to $1\,\rm{ms}$) as $t_{\rm obs}$. The resulting upper limits are 9\,$\upmu$Jy for the $T_{\rm{asc}}$ search, 12-30\,$\upmu$Jy for the blind accelerated searches over 1-hr to 5-minute segments, and 0.14\,Jy for the single-pulse searches, assuming a 10\% duty cycle for the pulses.

\section{Discussion and conclusions}
In this work, we performed a detailed analysis of the optical pulsations of the AMSP SAX~J1808 during its 2022 outburst using TNG/SiFAP2 observations. In addition, we searched for optical and radio pulsations during its quiescent state using simultaneous SiFAP2 and MeerKAT observations.

\subsection{Optical pulsations during outbursts} 
Optical pulsations were first discovered in 2019 during the initial rising phase of the outburst, when the source had a 0.5--10~keV X-ray flux of $F_{X,2019}\simeq3.1\times 10^{-11}\, \mathrm{erg\,cm^{-2}\,s^{-1}}$ \citep{Ambrosino_2021}. They were characterized by an optical pulse amplitude of $A^{\text{rms}}_{\text{2019}} = (0.55\pm 0.06)\%$ and a pulsed V-band optical luminosity of $L_{\rm{pulsed}}^{\rm V,2019} = (3.2\pm0.4) \times 10^{30} \,\text{erg s}^{-1}$ \citep{Ambrosino_2021}\footnote{Note that they adopted a broader wavelength range ($325-690\,\rm{nm}$) compared to our V-band and a distance of 3.5~kpc rather than 2.7~kpc \citep{galloway_2006, galloway_2024}.}. 

In this work, we extended the study of optical pulsations to the decay phase of the 2022 outburst.
We detected significant optical pulsations during both TNG/SiFAP2 observations. The X-ray flux was $F_{\rm{X1}}\simeq 4 \times 10^{-10}\, \mathrm{erg\,cm^{-2}\,s^{-1}}$ during the first night, approximately a factor of 2 higher than that on the second night. The corresponding pulsed optical luminosities were $L_{\rm{pulsed}}^{\rm V,O1} = (1.15\pm0.13) \times 10^{30} \,\text{erg s}^{-1}$ and $L_{\rm{pulsed}}^{\rm V,O2} = (1.38\pm0.13) \times 10^{30} \,\text{erg s}^{-1}$, respectively. 
This allowed us to investigate the properties of the optical pulsations over more than a decade in X-ray flux.

\begin{figure}
\centering
\resizebox{\hsize}{!}{\includegraphics[clip=true]{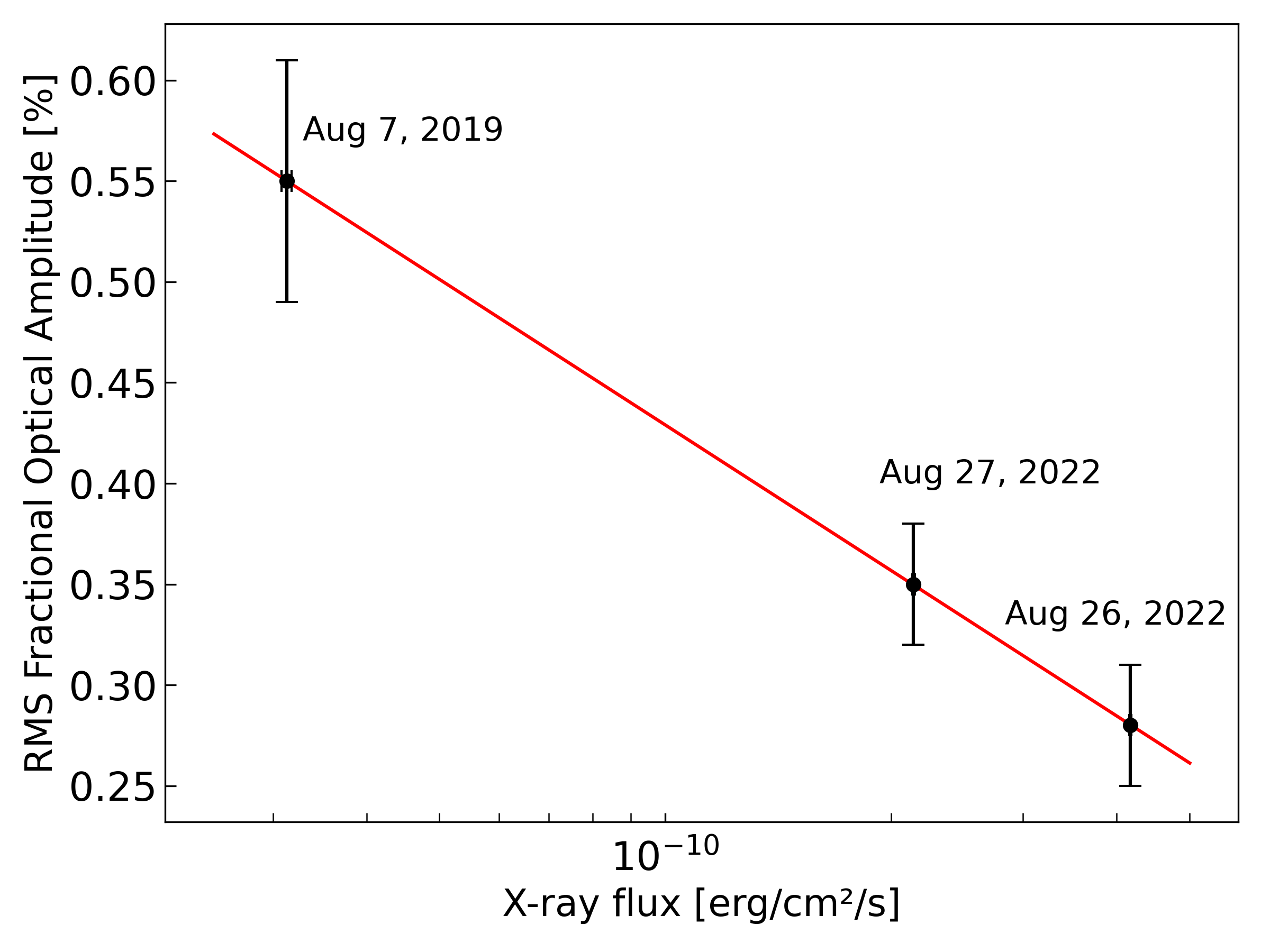}}
\caption{Fractional amplitude of the optical pulsations in O1 and O2, and during the 2019 outburst, as a function of the 0.5--10~keV X-ray flux derived from simultaneous NICER observations. The red line represents the overall trend.} 
\label{fig:ampvsflux}
\end{figure}

\begin{figure}
\centering
\resizebox{\hsize}{!}{\includegraphics[clip=true]{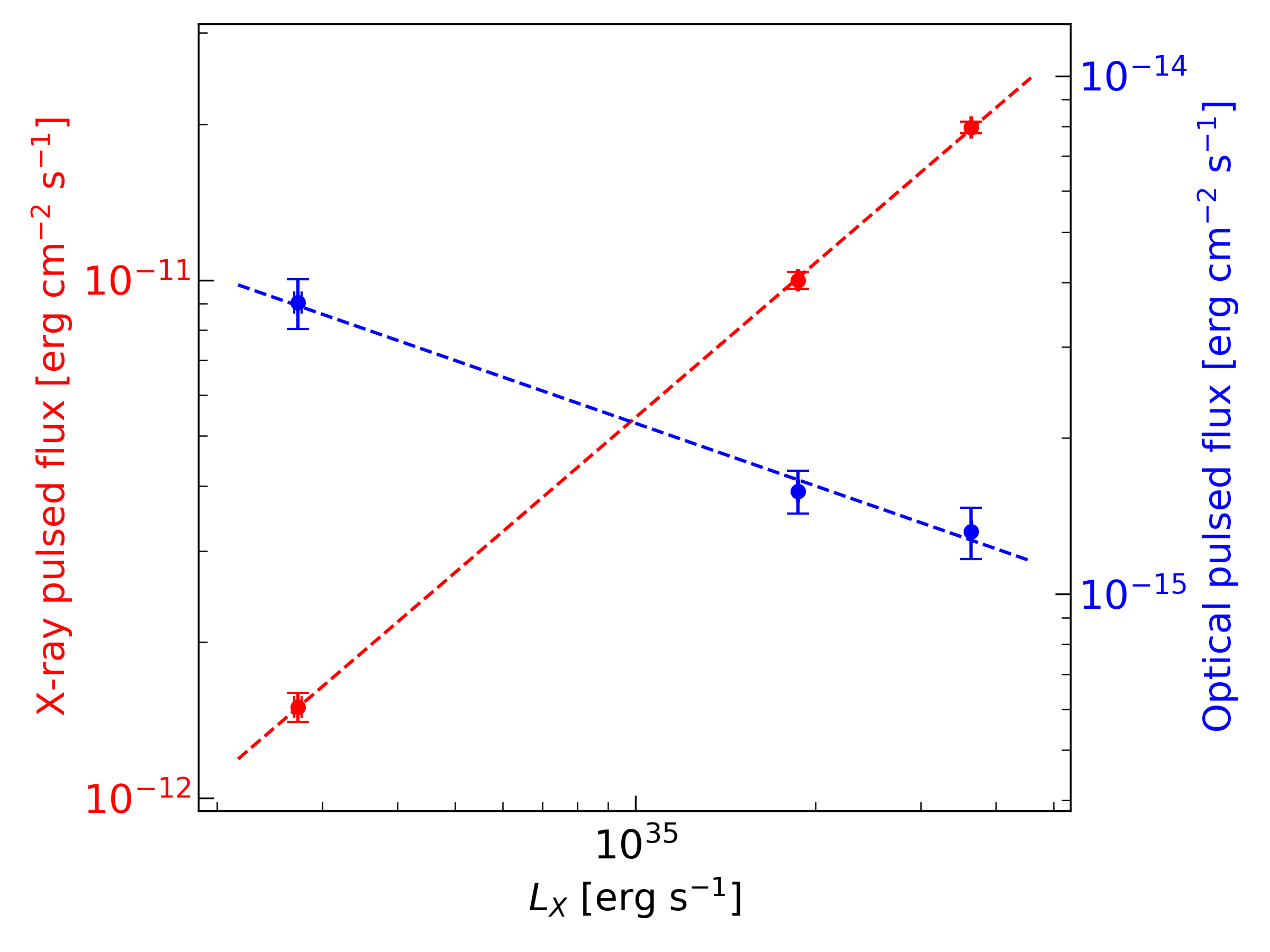}}
\caption{0.5--10~keV X-ray (red symbols, left axis) and optical (blue symbols, right axis) pulsed flux as a function of the X-ray luminosity $L_{\rm{X}}$, adopted as a proxy for the mass accretion rate. The dashed lines show the best-fit trends to the data.} 
\label{fig:anticorr}
\end{figure}

The fractional optical amplitude in the 2019 and 2022 observations decreased from about 0.55\% to 0.28\% with increasing X-ray flux (see Fig.~\ref{fig:ampvsflux}). Despite the limited number of measurements, this suggests a possible anti-correlation between the optical pulse amplitude and the X-ray flux. 
Assuming that the latter traces the instantaneous mass accretion rate $\dot{M}$, we compare in Fig.~\ref{fig:anticorr} the optical (blue points) and X-ray (red points) pulsed fluxes.   
The optical pulsed flux decreases with increasing X-ray luminosity (power-law index of about -0.4). In contrast, the X-ray pulsed flux scales approximately linearly with it, and exhibits a larger variability amplitude (factor of $\sim13$, compared to $\sim2.8$ in the optical band). This shows the different behavior of the optical pulsations with respect to the X-ray pulsations. 


\begin{figure}
\centering
\resizebox{\hsize}{!}{\includegraphics[clip=true]{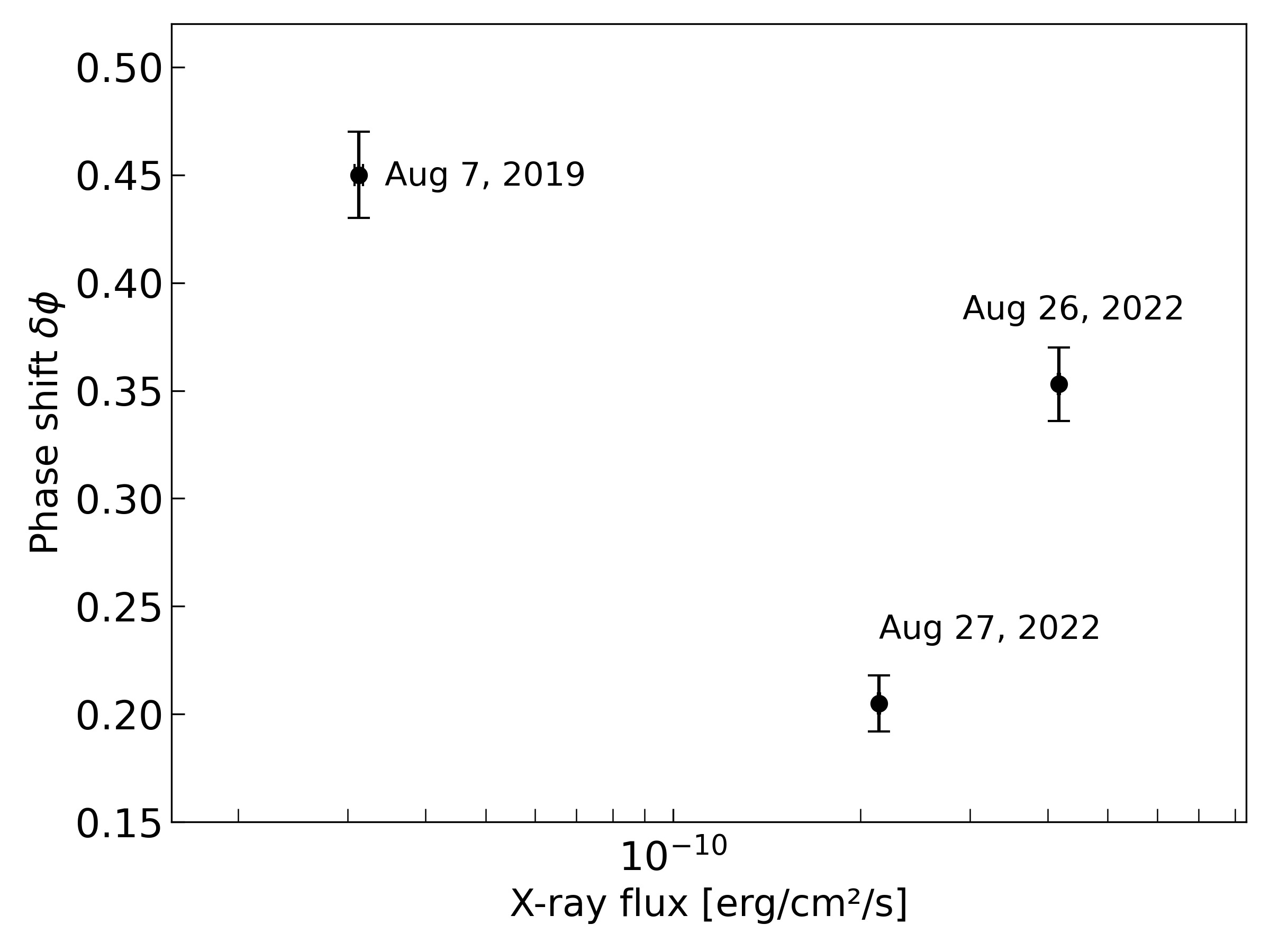}}
\caption{Phase shifts between X-ray and optical pulsations in SAX~J1808 in O1 and O2, and during the 2019 outburst, as a function of the 0.5--10~keV X-ray flux derived from simultaneous NICER observations.} 
\label{fig:phaseshiftsvsflux}
\end{figure}

The optical pulsations were found to be delayed relative to those in the X-rays. The measured phase shift varied from $\delta \phi_{2019} = 0.45\pm 0.02$ in the 2019 observation \citep{Ambrosino_2021} to $\delta \phi_{O1} = 0.353\pm0.017$ and $\delta \phi_{O2} = 0.205 \pm 0.013$ during 2022 observations (see Fig.~\ref{fig:lags}). 
These values (and the corresponding time lags) did not display a monotonic trend relative to the X-ray luminosity (see Fig.~\ref{fig:phaseshiftsvsflux}). However, they may be consistent with a decreasing behavior during the outburst. 
The simultaneous optical-to-X-ray SED of SAX~J1808 obtained in this work (Fig.~\ref{fig:sed}) showed that a single power-law component cannot account for the pulsed optical and X-ray fluxes. 
This is in agreement with the results reported in Fig.~12 of \citet{Ballocco_2026}, where the simultaneous X-ray and UV SED of SAX~J1808 similarly ruled out a single power-law.

The combined properties of the optical pulsations observed during the 2019 and 2022 outbursts emphasize that their origin cannot be explained in the framework of standard emission mechanisms in accreting millisecond X-ray pulsars. The simultaneous detection of X-ray pulsations shows that the X-ray emission is powered by magnetically channeled accretion onto the NS magnetic poles. In contrast, the optical pulsed luminosity is more than three orders of magnitude larger than that expected from thermal emission from the hotspots on the NS surface. At the same time, reprocessing of the high-energy emission in the outer disk cannot explain either the coherence or the flux of the optical pulsations \citep{Ambrosino_2021}. 

The markedly different luminosity dependence of the optical and X-ray pulsations, together with the observed phase shifts and the broadband SED, suggests that the two components originate from physically distinct emission regions. In particular, the anti-correlation between optical pulsed flux and X-ray luminosity indicates that the conditions responsible for producing the optical pulsations become more favorable at lower accretion rates. This behavior may be related to changes in the disc–magnetosphere interaction region as the inner disc recedes from the NS, which could provide more favorable conditions for particle acceleration and the production of optical and UV pulsations.

Interestingly, in a scenario similar to that proposed for PSR J1023+0038, where the optical pulsations arise from synchrotron emission powered by a pulsar wind interacting with the inner accretion disc \citep{Papitto_2019, Veledina_2019}, the intercepted wind power would scale approximately as $R_{\rm m}^2$, where $R_{\rm m}$ is the inner disk radius. Assuming the Alfvèn expression $R_{\rm m}\propto\dot{M}^{-2/7}$, this would imply $L_{\rm pulsed}^{\rm opt}\propto\dot{M}^{-4/7}$, broadly consistent with the observed anti-correlation ($L_{\rm pulsed}^{\rm opt}\propto L_{\rm X}^{-0.4}\propto \dot{M}^{-0.4}$, see Fig.~\ref{fig:anticorr}). However, this scenario requires the NS to launch a relativistic wind as a rotation-powered pulsar, and does not seem readily applicable to SAX~J1808, where simultaneous X-ray pulsations indicate that the accretion flow enters the magnetosphere. 

A possible interpretation is that the bright optical pulsed emission is generated via synchrotron radiation from relativistic particles accelerated in the interaction region between the accretion disc and the pulsar magnetosphere. In this framework, the optical pulsations are generated by a magnetospheric process that can coexist with the accretion-powered X-ray pulsations, naturally explaining their simultaneous presence while accounting for their different luminosity dependence and phase behavior. Overall, the observations of SAX~J1808 during the 2019 and 2022 outbursts support a scenario in which the optical pulsations trace a distinct emission component associated with the disc–magnetosphere interaction rather than a direct counterpart of the X-ray pulsations.

A useful comparison can be made with the tMSP PSR~J1023+0038, for which optical pulsations were first detected while the source was in an X-ray sub-luminous disk state with an average X-ray luminosity of $L_{\rm{X}} \simeq 4\times 10^{33}\, \text{erg s}^{-1}$ \citep{Ambrosino_2017}. In contrast to SAX~J1808, the optical and X-ray pulsations in this system are nearly phase-aligned and are predominantly detected during the so-called “high” X-ray mode (the source spends $\sim 70\%$ of the time in it), while they disappear in the low mode, suggesting a tight connection between the two energy bands \citep{Papitto_2019, Illiano_2023b}.
Moreover, the optical and X-ray pulse profiles show similar shapes, and their SED is well described by a single power-law extending from the optical to the X-ray band \citep{Papitto_2019}. These properties support the interpretation that both optical and X-ray pulsations in PSR~J1023+0038 originate from a common emission mechanism, most likely synchrotron radiation produced in a boundary region beyond the light-cylinder radius, where the relativistic pulsar wind interacts with the surrounding accretion flow \citep{Papitto_2019, Veledina_2019}. By contrast, the different luminosity dependence of the optical and X-ray pulsations, the variable phase shifts, and the optical-to-X-ray SED of SAX~J1808 indicate that its optical pulsations arise from a physically distinct emission component.

\subsection{Optical and radio pulsation search during quiescence}
\begin{figure}
\centering
\resizebox{\hsize}{!}{\includegraphics[clip=true]{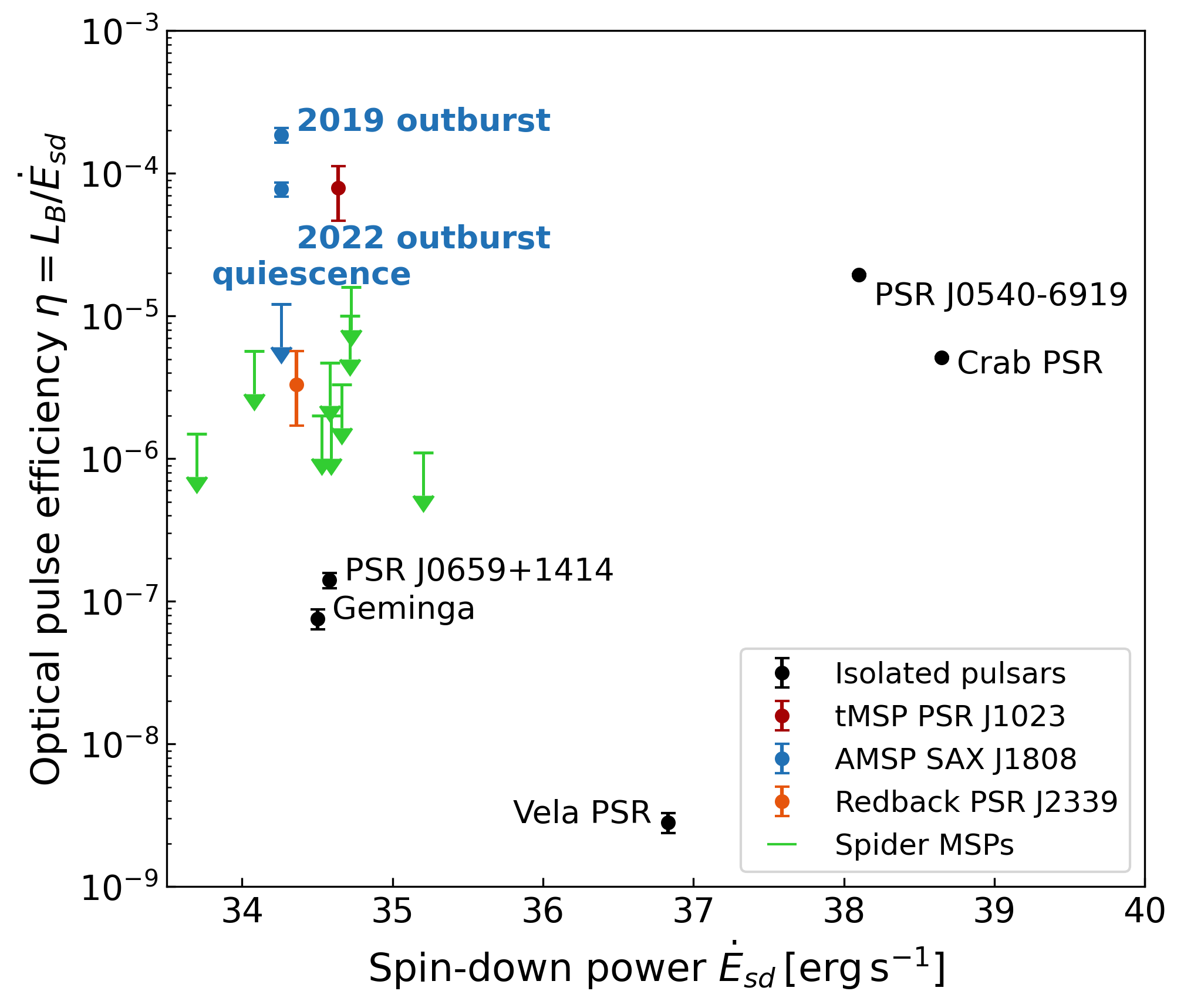}}
\caption{Optical pulse efficiency as a function of the spin-down power for SAX~J1808 in outburst and quiescence, compared with other known optical pulsars. The optical pulse efficiency is defined as the ratio of the pulsed B ($\lambda_{\text{eff}}= 433\, \text{nm}$, $\text{W}_{\text{eff}}= 90\, \text{nm}$) luminosity to the spin-down power. The x-axis represents the base-10 logarithm of the spin-down power values.
This figure is an updated version of the one presented in \citet{Malacaria_2026}, and previously in \citet{Papitto_2025} (see also \citealt{Ambrosino_2017} and references therein). The downward arrows represent the 95\% c.l. upper limits on the optical pulse efficiency. Error bars show uncertainties at the $1\sigma$ confidence level. 
} 
\label{fig:efficency_plot}
\end{figure}

In the optical observations obtained during the quiescent state of SAX~J1808, we found no significant pulsation signal. We could thus place a 95\% c.l. upper limit on the pulsed optical quiescent emission of $F^{\text{V, UL}}_{\text{pulsed}}<2.5\times 10^{-16}\, \mathrm{erg\,cm^{-2}\,s^{-1}}$. 
The resulting upper limit on the optical pulsed efficiency $\eta$, defined as the ratio of the pulsed optical luminosity to the spin-down power, is computed in the B band\footnote{The B-band upper limit was obtained from the V-band measurement assuming a flat spectral flux density between the V and B bands.} ($\lambda_{\text{eff}}=433~\text{nm}$ and $\mathrm{W}_{\text{eff}}=90~\text{nm}$) to allow a consistent comparison with previous measurements of other known optical pulsars reported in the literature (see Fig.~\ref{fig:efficency_plot}).
In particular, isolated and young rotation-powered pulsars exhibit optical pulse efficiencies in the range  $\eta \approx 10^{-9} - 10^{-5}$ (see black dots in Fig.~\ref{fig:efficency_plot}). The optical emission from these pulsars is generally interpreted as incoherent synchrotron radiation produced by relativistic particles accelerated within the NS magnetosphere \citep{Pacini_1983}.
As found for the redback radio-pulsar PSR~J2339-0533 \citep{Papitto_2025}, the optical pulsed efficiency of SAX~J1808 during quiescence could be comparable to that of the Crab pulsar, despite being more than four orders of magnitude less energetic in spin-down power.
The upper limit obtained in this work for the quiescent state of SAX~J1808 ($\eta < 1.2\times 10^{-5}$) is also consistent with those derived by \citet{Malacaria_2026} for a sample of eight spider radio MSPs observed with SiFAP2 in search for optical pulsed signals (see green arrows in Fig.~\ref{fig:efficency_plot}).
This finding does not disfavor the possibility that a common mechanism underlies the production of optical pulsations in these systems. We thus conclude that the upper limit on the optical pulsed amplitude during quiescence does not rule out the presence of an active rotation-powered pulsar during the quiescent state of SAX~J1808.

To further test this hypothesis, (quasi)-simultaneous observations were carried out in quiescence also in the radio band with the MeerKAT radio telescope. No pulsations were detected down to $\sim10$\,$\upmu$Jy, a value that encompasses Galactic-field MSPs whose 1.4\,GHz flux densities are listed in the ATNF pulsar catalogue\footnote{\url{https://www.atnf.csiro.au/research/pulsar/psrcat/}} \citep{psrcat}. This value is at least $\sim5$ times (depending on the assumed spectral index) smaller than the previous deepest upper limits obtained for this source \citep{Patruno_2017}, and among the most stringent upper limits on radio pulsations for any AMSP (see \citealt{Marino_2026} for a summary).
A search for single dispersed pulses also resulted in non-detections down to a peak-flux limit of 0.14~Jy. Given the relatively low observing frequency used in this search, and assuming the radio beam intersects our line of sight, the most likely reason for our non-detection is absorption of the radio signal by intra-binary matter enshrouding the system, including during quiescence. 
Observations with the MeerKAT S-band receiver \citep[unavailable at the time of observing]{2018IAUS..337..175B} and with the future band-5B receiver \citep{band5} will allow us to test this interpretation further by going deeper at higher frequencies less obstructed by the medium, and possibly reveal SAX~J1808's radio pulsations. 

The significant decrease in the optical pulsed flux of SAX~J1808 observed during quiescence (more than an order of magnitude compared to the outburst state, similar to what is observed in redback systems) further suggests that the optical emission mechanism becomes substantially more efficient in the presence of an active accretion flow, possibly due to enhanced particle acceleration at the interface between the accretion disc and the NS magnetosphere.
In a broader context, the tMSP PSR~J1023+0038 and SAX~J1808 during its outburst state exhibit optical pulsation efficiencies that are approximately $50$–$100$ times higher than those observed from the other known optical pulsars, suggesting that the presence of an accretion disc around the NS may play a key role in enhancing particle acceleration processes and the resulting synchrotron emission \citep{Papitto_2019, Veledina_2019, Romei_2026}.

Future observations spanning a broader range of mass accretion rates will be crucial for testing this scenario and establishing how the efficiency of the optical pulsation mechanism evolves across different accretion regimes.

\begin{acknowledgements}
This work was supported by INAF (Research Grants ``Uncovering the optical beat of the fastest magnetized neutron stars (FANS)'', ``Polarized X-rays from an accreting millisecond pulsar: a pathway to the equation of state of neutron stars (PULSE-X)'' (PI: Papitto), ``iPeSKA'' (PI: Possenti), and ``Chasing the optical counterpart in FRBs and Magnetars with SiFAP2@TNG and SiFAP-SoFT@Cassini`` (PI: Ambrosino)); by the Italian Ministry of University and Research (MUR PRIN 2020) Grant 2020BRP57Z, ``Gravitational and Electromagnetic-wave Sources in the Universe with current and next-generation detectors (GEMS)'' (PI: Astone); by Fondazione Cariplo/Cassa Depositi e Prestiti through Grant 2023-2560, ``Taking the optical pulse of the quickest spinning Neutron Stars: a pilot exploratory study (SPES)'' (PI: Papitto); and in part by the Italian Ministry of Foreign Affairs and International Cooperation under Grant ZA23GR03 for the project ``RADIOMAP -- Science and technology pathways to MeerKAT+: the Italian and South African synergy''. GI is supported by a Juan de la Cierva fellowship (JDC2024-053550-I). MCB acknowledges support from the INAF Fundamental Research Grant XBOOM. AM acknowledges support from the Fund Vera Rubin/Chile 2024, under the project DIA 1736 "Silent black holes around red supergiants". DMR is supported by Tamkeen under the NYU Abu Dhabi Research Institute grant CASS. FCZ is supported by a Ramón y Cajal fellowship (grant agreement RYC2021-030888-I). FCZ and AM acknowledge support from the Spanish grant PID2023-153099NA-I00 and from the Unidad de Excelencia María de Maeztu programme CEX2020-001058-M.

The MeerKAT telescope is operated by the South African Radio Astronomy Observatory (SARAO), a facility of the National Research Foundation, an agency of the Department of Science and Innovation. We thank the SARAO staff for their support with observations and commissioning. The MeerKAT observations made use of the FBFUSE and APSUSE computing clusters for data acquisition, storage, and analysis. These clusters were funded and installed by the Max-Planck-Institut f\"ur Radioastronomie (MPIfR) and the Max-Planck-Gesellschaft.

This work also makes use of observations from the Las Cumbres Observatory global telescope network. The authors thank Daniel Bramich for developing the XB-NEWS pipeline, which was used in this work.
\end{acknowledgements}

\bibliographystyle{aa} 
\bibliography{refs}
\end{document}